\documentclass{article}
\usepackage{graphicx} 
\usepackage{amsmath,amssymb}
\usepackage{cite}
\usepackage{nameref,hyperref}
\usepackage[table]{xcolor}

\title{How high-resolution agent-based models can improve fundamental insights in tissue development and cell culturing methods}
\author{Paul Van Liedekerke$^1$, Jiří Pešek$^2$, Kevin Alessandri$^3$, Dirk Drasdo$^2$  }
\date{%
    $^1$BionamiX group, Department Data Analysis and Mathematical Modelling, Ghent University, Gent, Belgium\\%
    $^2$National Institute for Research in Computer Science and Automation (INRIA de Saclay), Palaiseau, France\\[2ex]%
    $^3$Treefrog Therapeutics, F-33600, Pessac, France\\[3ex]%
    \today
}

\begin{document}

\maketitle

\section*{Abstract}

The fundamental understanding of how cells physically interact with each other and their environment is key to understanding their organisation in living tissues. Over the past decades several computational methods have been developed to decipher emergent multi-cellular behaviors. In particular agent-based (or cell-based) models that consider the individual cell as basic modeling unit tracked in space and time enjoy increasing interest across scientific communities. In this article we explore a particular class of cell-based models, so-called Deformable Cell Models (DCMs), that allow to simulate the biophysics of the cell with high realism. After situating this model among other model types, We give an overview of past and recent DCM developments and discuss new simulation results of several applications covering in-vitro and in-vivo systems. Our goal is to demonstrate how such models can generate quantitative added value in biological and biotechnological problems.

\section{Introduction}

Living tissues can be considered as active soft matter \cite{Marchetti2013,Shaebani2020}. Besides the genetics and many molecular pathways that regulate metabolism and growth of the organisms, physical forces (mechanical stresses) are known to interfere strongly with these mechanisms in either a passive or active way \cite{Trepat2009,Di2023,Hunter2025}. Because of this complexity, mathematical and computational modeling approaches have become key in understanding this interplay, attempting to combine physical and biological mechanisms to decipher the spatial organization of cells in biological tissues \cite{Sunyer2016,Metzcar2019}.

The methods and approaches to describe the spatio-temporal dynamics of tissues can be broadly divided into two classes\cite{DAlessandro2015,Byrne2009}.

On the one side, continuum tissue models regard biological tissues as a continuous matter, discarding its "granular" structure made of cells. Such approaches describe quantities ("fields") that  reflect local averages over a sufficiently large cell numbers. These rely on the well-known continuum equations for momentum, energy and mass (cell density) balances (e.g.cite{\cite{Lowengrub2009,Frieboes2006}). Their ability to perform rigorous mathematical analysis and their computational efficiency  due to the scale invariance of the equations has made these methods widespread. Although continuum methods do not capture the inherent variability nor resolve the spatial scale of individual cells, they may capture the dynamics at the level of cellular subpopulations \cite{Perthame2018}.
On the other hand, Agent-based models (ABM) treat the tissue naturally as a discrete system of elements, using the natural alphabet of cells and tissues. In an ABM, every cell is discerned along with its own physical properties, and its phenotypic and genotypic variability and behavior. The ability of capturing the motion of each cell, the individual cell variability, and emergent behavior of many interacting cells is one key strength of the ABM.
Research in both of the above mentioned approaches have resulted in hybrid methods \cite{Kim2007} whereas mathematical foundations linking the equations in ABMs to those of continuum models can give insights in the parameter relations between them (e.g. \cite{Chaplain2020, Ardaševa2020}).

ABMs can be further classified into "lattice-based" models, where cell movement is restricted to a space-fixed lattice \cite{Drasdo2018}, and "lattice-free" models, where cells can continuously  move in space \cite{VanLiedekerke2018}. This work considers the latter. One of the first lattice-free ABMs for cells dates back to the '90, in which each cell was approximated as a circular object, thought to represent the space where most of the cell volume is located \cite{Drasdo1995}. 
This is the prototype of a "center-based model" (CBM, see Fig. \ref{fig:all_models}A) where interaction energies for cells have been mimicked by interaction potentials  between cell centers. 
Because of the flexibility and extensibility of the approach, most recent CBMs usually mimic the change of the cell center position $\vec{x}_i$ with time by a overdamped force-balance equation (equation of motion), which in its basic form reads:
\begin{equation}\label{eq:CBM_motion}
\Gamma_{cm} \vec{v}_i  + \sum_{j}\Gamma_{cc}(\vec{v}_i-\vec{v}_j) = \sum_{j}\vec{F}_{cc,ij} +  \vec{F}_{mig,i},  
\end{equation}
with $\vec{v}_i=d\vec{x}_i/dt$, $i=1, ...,N$ enumerating all $N$ cells.
Here, the terms on the lhs. denote cell-medium and cell-cell friction, respectively, with $\Gamma_{cm}$ being the cell-medium friction tensor, $\Gamma_{cc}$ being the cell-cell friction tensor. This friction accounts for viscous shear resistance in cell-cell contacts. Several authors tend to ignore cell-cell friction leading to de-coupling of the cell velocities and a significant computational speedup, but this inherently means that shear viscous forces between cells are neglected. In principle such simplifications should be verified beforehand to have an estimate of the impact on the simulation results. For example, in simulations of cell migration within dense aggregates, cell-cell friction may play an significant role.

The cells in a CBM mechanically interact through a contact force $\vec{F}_{cc,ij}$ that depends on the distance between the cell centers, polarizations, or other attributes. This contact is often treated as a two-body contact, and the force law may be purely phenomenological or based on certain cell material assumptions. Example for the latter are Hertz's contact force and the JKR model that approximates a cell by an isotropic homogeneous elastic sphere, and is valid for small deformations. The fact that in a multi-cellular environment, force relations derived for two-body contacts cannot take into account the simultaneous effect of many interaction forces by neighboring cells is frequently ignored, possibly leading to inaccurate results in particular in case of highly compressed cells.\cite{VanLiedekerke2019}. The force $\vec{F}_{mig,i}$ describes the propulsion needed for the cell to overcome friction and migrate. The latter usually is assumed to have a stochastic components  mimicking the random micro-motility.  Note that other forces may be added to the equation depending on the system.

Historically, despite cell deformation has often been neglected in models, several studies highlight the role of cell shape and deformation in the division process and tissue morphology (e.g. \cite{Lovegrove2025,Ong2025}). Although the concept of a CBM can be extended to permit more complex cell shapes, for example ellipsoid-like  or rod-like shapes \cite{Palsson2007,Marin-Riera2019}, these concepts a priori only apply to systems in which an "average" cell shape prevails (e.g. bacteria). 
In systems where cells show complex patterns and large variation of cell shapes occur, the average cell shape hypothesis may not apply (see section \ref{sec:Applications}), and models with an intrinsic higher  resolution allowing an realistic description of cell shape are required. In the next section we will briefly describe a few actual modeling approaches that explicitly represent variable cell shapes, namely: Phase-field models (PFM), Cellular Potts models (CPM), Vertex models (VM), the Subcellular Element Method (SEM), and the Deformable Cell Models (DCMs) (Fig. \ref{fig:all_models}B-E). This paper summarizes the recent developments of the latter one.

\begin{figure}[h!]
    \centering
    \includegraphics[width=0.9\linewidth]{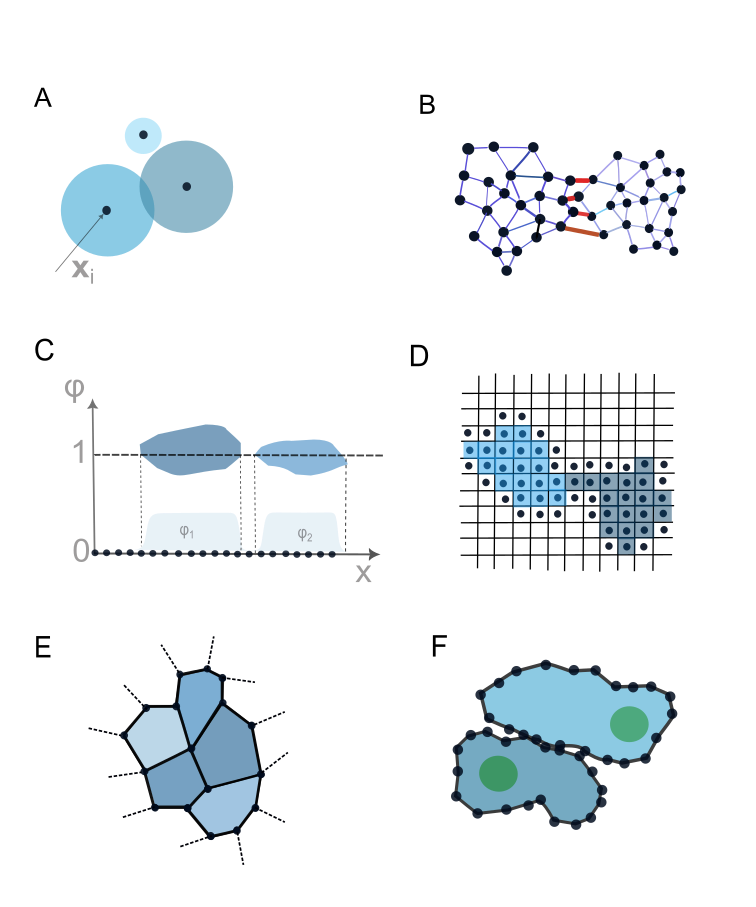}
    \caption{2D illustration of representations of frequently used cell-based models. In all cartoons the small black filled circles denote computational nodes for each model. A: center-based model (CBM). Cells are represented by spheres, B: Subcellular Element Model (SEM). Two cells are in contact. The lines between the nodes indicate the influence of an intracellular (blue) or intercellular (red) force potential, C: Phase-field model (PFM) for two 2D cells with phases given for 1D, D: Cellular Potts Model (CPM) for two cells, E: Vertex model (VM) for epithelial cells, F: Deformable Cell Model (DCM) for two cells. }
    \label{fig:all_models}
\end{figure}

\section{High-resolution cell model types}

We first give a short state-of-the-art for what one could categorize as "high resolution" cell models. There are several approaches in which cell models can incorporate biophysical detail. Basically  all address complex cell shape by increasing the number of degrees of freedom describing the geometry of the cell, but among them there are methodological differences as well as similarities, which can sometimes lead to name confusion in literature. Here, we will try to make a clear distinction based on some key elements.

 \paragraph{Subcellular Element Model (SEM):} Models with higher degrees of freedom can be obtained from extending the classic CBM having a unique center, to a model with multiple interacting centers. In the original model of Newman et al.  \cite{Newman2005}, called the Subcellular Element Model,  cells are mimicked by a set of particles that interact through pair-wise potentials resembling those used in molecular dynamics simulations (Fig. \ref{fig:all_models}B). These potentials are shaped such that when two nodes are closing in, a repelling force develops; whereas when they try to separate, an attractive force develops,  allowing to mimic the cohesive viscoelastic structure in a cell and adhesion between two cells.  The governing equations of motion of those particles are posed as \cite{Chattaraj2023}:
\begin{equation}
\label{eq:motion_VM}
\mu \frac{d\vec{x}_i}{dt} = \vec{F}_{R,i} + \sum_j\vec{F}_{C,ij}, 
\end{equation}
where $\vec{x_i}$ is the particle position, $\vec{F_{R,i}}$ is a random fluctuating force and $\vec{F_{C,ij}}$ is an interaction force between particle $i$ and $j$, derived from a soft potential function described by an interaction strength and cut-off distance for which beyond the interaction vanishes. The same potential function form can describe both intracellular and intercellular (cell-cell adhesion) interactions.  
This model is relatively straightforward to implement but it has a few critical weaknesses. The pairwise potentials between the nodes of the cells account that for the mechanical behavior (Fig. \ref{fig:all_models}B, blue connections), are coarse-grained versions of Lennard-Jones or Morse potentials and their parameters  (strength, interaction range) must be tuned according to the prescribed number of nodes per cell. The same potentials are used to model the cell-cell adhesion, mimicking the cadherin bonds forces (see Fig. \ref{fig:all_models}B, red connections). It is not always clear how the  parameters can be mapped to particular physical (macroscopic) properties of the cell, or how they could be obtained through experiments such as optical stretchers, tweezers, micro-pipettes or atomic force microscopy experiments \cite{Chattaraj2023}. 
The original SEM does not clearly discriminate between the nodes of the cortex and those of the inner parts. In particular when an insufficient number of nodes per cell are used, high forces between cells can cause un-physical interpenetration of cells when vertices of one cell pass those of the other, resulting in  phantom interactions.  The model does not explicitly account for viscous shear effects as the nodes only feel the viscous drag with a static medium. Frictional forces arising from nodes moving past each other may emerge in case of adhesive bond breaking an re-establishment events, but it is unclear how this friction can be parametrized in the model. Finally, we note that the SEM discretizes the whole cell volume with particles, hence computational times can raise quickly if high resolution is required.

\paragraph{Phase-field Model (PFM):} Phase-field models have originally been conceived in material sciences as mathematical models for solving interface problems, and have enjoyed an increased interest in cell mechanics in recent years. In phase-field models, individual cells are assumed to be separated continua, and their boundaries are described by a phase function $\phi(\vec{x},t)$ that can acquire values between 0 and 1 (for example, with with $\phi=0$ being the outside of the cell and $\phi=1$ inside of the cell, see Fig. \ref{fig:all_models}C) \cite{Kuang2003,Wenzel2021,Nielsen2020,Moure2021,Chiang2024}. The evolution of the phase is governed by a mass balance partial differential equation (similar to the Cahn-Hilliard type describing phase separations) that can be written in a  compact form for a cell $i$ as:
\begin{equation}
\label{eq:phase_field}
\frac{\partial \phi_i}{\partial t} + \vec{v}_i\cdot \nabla_{x_i} \phi_i = -\frac{\delta\mathcal{F}}{\delta \phi_i} 
\end{equation}
Here $\mathcal{F}$ is a free energy functional that describes the mechanical state of the cell as well as the contact with other cells. It can include components that account for cell shape changes, cell polarization, and more. Friction between cells can be captured as well \cite{Chiang2024}. Standard PFMs allow to incorporate well-known mechanics laws but require good understanding of continuum theories and the energy functionals may be non-trivial to parameterize. In particular, including mechanics of subcellular elements such as nucleus or cytoskeleton  will likely pose some challenges \cite{Monfared2025}.  The equations \ref{eq:phase_field} are solved on a fixed (Eulerian) background grid, which can become expensive in 3D for large cell numbers or large domains  \cite{Kuang2023}. The cell boundaries are not tracked explicitly and may not be arbitrarily sharp defined.

\paragraph{Cellular Potts model (CPMs):}  These are lattice-based models in which each cell occupies several lattice sites (Fig. \ref{fig:all_models}D). The CPM is in essence a pure stochastic method. The global system state is described by a set of parameters and state variables, while transitions between two states are controlled by a discrete form of an energy-like functional that accounts for energy contributions and changes arising from the individual cell properties and constraints (e.g. volume, surface, migration,..) and those from interactions among cells (e.g. adhesion) \cite{Glazier1993}. Probabilistic transitions from one system state to another occur according to a transition rate that itself is a function of the energy functional. 
In principle, the temporal evolution of the multivariate probability of finding the system in a certain state follows a master equation, but as the state space is very large, usually Metropolis sampling is applied. 
This method assumes that state changes of the multi-cellular system are driven by the attempt of the system to relax into equilibrium, which may distort the natural time scales present in the system. 
Time is usually "measured" by the number of Monte-Carlo steps (for a more exhaustive discussion, see ref. \cite{VanLiedekerke2015}). Overall, CMP are less suited for systems entailing events occurring at multiple timescales. Recently a so-called Poissonian CPM has been developed \cite{Belousov2024}. In this approach a master equation for the voxels is being solved that should allow to simulate an undistorted dynamics of physical phenomena. However, this may significantly increase computational times compared to the traditional CPM. CPMs are widespread, relatively easy to implement, and particularly strong in providing qualitative insights in pattern formation in tissues, including cell sorting and rearrangements, wound closure, and interactions of cells with extracellular matrix (e.g.\cite{Canty2017,Hirashima2017,Tsingos2023}).  On the other hand, certain localized or filamentous subcellular structures may be non-trial to incorporate. In 3D, the CPM may become  computationally too demanding because of the entire cell volume discretization \cite{Sultan2023}. Basic CPMs can also introduce lattice artifacts \cite{VanLiedekerke2015,Nemati2024}.

\paragraph{Vertex model (VM):} In VM cells are assumed to have a polyhedral shapes (polygons in 2D) formed by vertices (nodes) that can move continuously in space. As in this approach adjacent cells share common borders, they inherently mimic compact multicellular assemblies (Fig. \ref{fig:all_models}E) \cite{Fletcher2014}.  A cell's shape is determined by the forces on the corners of the borders that it shares with its neighboring cells. The forces include contributions from cell volume, cell surface, adhesion tension and cortical tension forces. The cell shape dynamics is governed by a force balance equation (equation of motion) that can be written compactly as:
\begin{align}
\label{eq:motion_VM}
\mu \vec{v_i} = -\nabla_{\vec{x_i}} \mathcal{E}, 
\end{align}
where $\vec{x_i}$ and $\vec{v_i}$ are the vertex position and velocity, $\mathcal{E}$ is the energy term that accounts for changes in cell volume, surface area, and surface tension, and $\mu$ is a friction term that controls the mechanical timescale of the system.
The VM is a very popular and particularly a good choice for epithelial tissue modeling in which the cells do not physically detach from each other (see e.g.\cite{Kong2016, Barton2017,Silvanus2017, Okuda2018, Ioannou2020,Sego2023, Brinas2024}). 
However, in the traditional VM approach, a cell does not exist as an independent, isolated entity, limiting its applicability to situations in which cells do not become separated or de-tach and (re)attach to each other. VMs are computationally quite efficient and can handle large cell numbers, even in 3D.

\paragraph{Deformable cell model (DCM):} The remainder of this paper is devoted to this class of models. In the DCM cells are clearly delimited by a surface triangulation (in 3D) for each cell individually that surrounds the cytoplasm and nucleus. The nodes forming these triangles can move in continuum space.  For each vertex (node) of the triangles a force balance equation is solved. Contact and interaction between triangles is dynamically updated every timestep. The forces arise from known viscoelastic laws, theories of adhesive contacts, and physical constraints (such as fluid incompressibility). 
A DCM permits description of complex cell shapes both for cells in isolation as for cells in multicellular assemblies (Fig. \ref{fig:all_models}F).
Early developments of this model type in 3D can be traced back to efforts to describe the dynamics of red blood cells\cite{Noguchi2005,Fedosov2010} and plant cells \cite{VanLiedekerke2010}. Several refinements and improvements have been done since then and those are still ongoing (e.g. \cite{Odenthal2013, Smeets2019, Bodenstein2020,VanLiedekerke2020, TorresSanchez2022,Okuda2023, Runser2024}).  Because the model uses a surface discretization, the computational efforts rise moderately with the number of cells compared to methods that apply a volume discretization. On the other hand, the method's performance depends also on the efficacy of triangle-triangle contact detection. Using Axis-Aligned Bounding Box (AABB) trees, this can be done efficiently. Alternatively researchers have also developed 2D variants \cite{Rejniak2007,Jamali2010, Tozluoğlu2013, Cammarota2024}. 


Standard SEMs are based on a volume discretization, whereas DCM involves a surface discretization and clearly discerns the cell boundary from the cell interior. This allows to explicitly model the contractile state of the cell cortex which generally has a large impact on tissue organization. Hereby a sufficient number of surface vertices is imposed (depending on the system size and detail) because cell cortex force models are derived from constitutive laws and a too coarse discretization can introduce significant artifacts. A DCM includes viscous cell-cell friction forces explicitly (see further), whereas in SEM cell-cell friction this is an emergent effect arising from the intercellular potentials between particles.
Compared to phase-field models, while a basic DCM only keeps track of the boundary vertices of the cell, localized and heterogeneous structures such as focal adhesions and stress fibers are likely to be incorporated more straightforwardly in the particle-based structure of a DCM. On the other hand phase-field models allow to naturally include various intracellular and extracellular fields (e.g. ECM density) in the same modelling framework, and in principle can make use of efficient software packages (e.g. Finite Element solvers).
Compared to the VM, the DCM naturally incorporates  (local) contact establishment and detachment between cells, whereas in a VM this is enforced by so-called T-transitions which changes the local connectivity of the vertices. The cells in a DCM have more degrees of freedom and shape is not prescribed (like the polyhedral shape in VM) but acquired  based on the force balance. Recent developments \cite{Kim2021} in VM allow more complex behavior in cell dynamics, including treatment of interstitial spaces and adaption to more complex shapes. These modifications can make the VM more realistic and tend to efface the differences between a DCM and VM, but are likely to be implemented ad hoc in dedicated VM softwares.

It is clear that all high resolution models generally come at the expense of a higher computational cost than a CBM, hence the spatial dimensions of the envisaged systems are more limited. The applicability of a DCM on a single workstation is often restricted to 1000 to 10,000 cells, depending on the desired accuracy and simulation time. As such they are merely suited for systems of that size (e.g. small bio-engineered in-vitro systems or systems in early embryonic development) or to sub-systems in larger tissues whereby the remaining parts of the tissue can be accounted for in a hybrid fashion by a coarser scale model or simpler ABM such as the CBM \cite{VanLiedekerke2022}. With the statements made above we do not pursue to make any value judgments. The question which model to use among the listed will depend on the biological system type and complexity, the timescales involved, the software availability and usability, and  the experience of the user or research group.

In the next section we will first recapitulate the basic model components of DCM. The following sections also showcase several examples for which the DCM has been used and we will discuss some frequently used bio-engineered systems for which the DCM can be deployed as a valuable research tool. We will illustrate this by reviewing several precedent works as well as providing new simulations.

\section{Overview of DCM components}

In this section we briefly summarize the common building blocks of the DCM. While some of these are necessary for basic functioning, others can be added to incorporate more phenomena depending on the hypotheses that should be tested. The exact implementation of those components can differ according to the preferred detail as well as according to the assumed mechanisms or hypotheses. 
For these reasons, we will provide a limited technical description, as we focus on a high-level overview of the underlying concepts. Further details can be found in the referenced literature. 

\subsection{Basic model structure}

Essentially, a DCM represents the cell as a shell that contains a homogeneous, isotropic fluid. In 3D, the cell surface is modeled by a triangulated surface mesh (see Fig.\ref{fig:cell_division}A). On the surface force fields are applied, arising from mechanical deformations of the cortex, external cell-cell interactions, cell migration, volumetric constraints (e.g. originating from volume conservation if the membrane is impermeable), as well osmotic effects. Due to the discrete nature of the surface mesh, the force fields are then integrated over the neighborhood of each node to represent a local force applied on a single node. In practice, this step is often done implicitly via distribution rules on e.g. contact forces. 
The dynamics of each node is described by a force balance equation for each node according to Newton's second law, whereby inertia effects are usually neglected as they are small compared to friction forces. This results in a overdamped Langevin equation.

\begin{figure}[h!]
    \centering
    \includegraphics[width=0.99\linewidth]{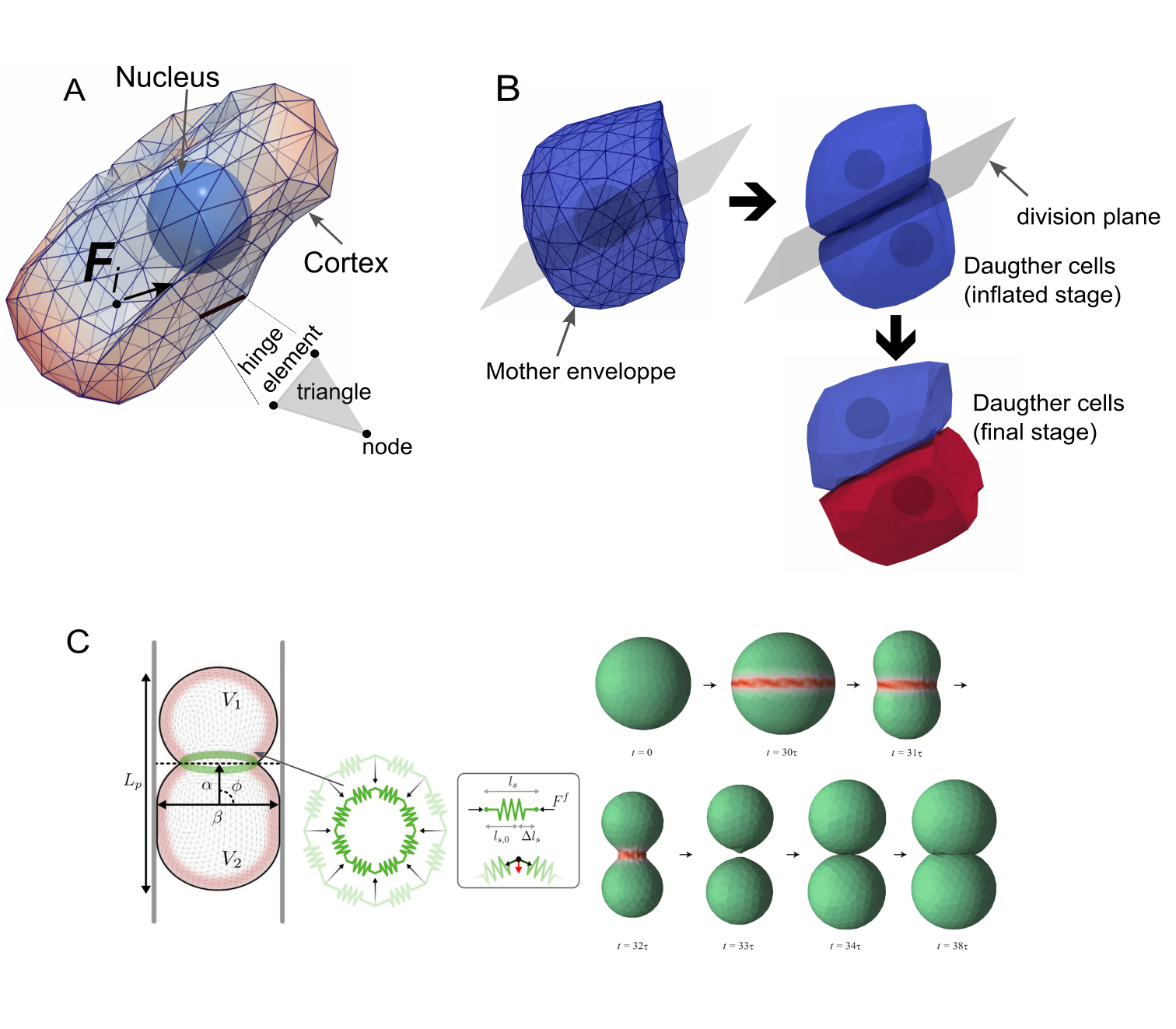}
    \caption{ A: DCM with its cortex triangulation and nucleus. B-C Two different approaches to simulate cell division in DCM. B: Inside the mother cell envelope, two new cells are created that are separated by the division plane. These daughter cells inflate until they touch the mother envelope. The latter is finally removed. Re-meshing is not necessary. C: The cytokinesis process is simulated, with a furrow contraction of the cortex along the division plane. Two daughter cells emerge when the contractile ring is small enough. In this case, re-meshing is necessary (picture taken from \cite{Cuvelier2023}).}
    \label{fig:cell_division}
\end{figure}

For any node $i$ of a cell\footnote{The cell index has been dropped here for clarity.} in which $\vec{v}_i$ denotes its velocity (see Fig.~\ref{fig:cell_division}A), the force balance equation (equation of motion) reads:
\begin{multline}
\label{eq:motion_DCM}
\underbrace{\Gamma^{}_{ns,i} \vec{v}_i}_{\textrm{substrate/ECM friction}}
+\underbrace{\sum_{j}\Gamma^{}_{nn,ij} (\vec{v}_i-\vec{v}_j)}_{\textrm{node-node friction}}  
=  \underbrace{\sum_{j}\vec{F}^{}_{e,ij}}_{\textrm{in-plane el. force}} 
+ \underbrace{\sum_{m}\vec{F}^{}_{m,i }}_{\textrm{bending force}} 
 \\ 
+ \underbrace{\sum_{j}\vec{F}^{}_{s,ij}}_{\textrm{membrane tension}}  
+ \underbrace{\vec{F}^{}_{vol,i}}_{\textrm{volume changes}} 
+ \underbrace{\vec{F}^{}_{osm,i}}_{\textrm{osmosis}}
\\
+ \underbrace{\vec{F}^{}_{rep,i} + \vec{F}^{}_{adh,i}}_{\textrm{contact}}
+ \underbrace{\vec{F}^{}_{fluc,i}}_{\textrm{fluctuations}}
+ \underbrace{\vec{F}^{}_{mig,i}}_{\textrm{migration}} .
\end{multline}
Here $\Gamma_{ns}$ is a matrix that represents the friction effects due to cell-substrate or cell-extracellular matrix (ECM) interactions. The substrate (e.g. the ECM) is usually modeled implicitly as a field quantity which itself is not affected by the cells (in this case the total momentum of the whole system is not conserved). A recent study on fibrosis however, treats the influence of ECM fibers explicitly \cite{Zhao2023}. The second matrix $\Gamma_{nn}$ is the node-node friction matrix originating from intracellular viscous effects or friction between two adjacent cells. 
Friction is often assumed to be isotropic and determined by a scalar coefficient $\gamma$ (see Table \ref{tab:parameters}). Some of these values can be deduced from tissue viscosity measurements \cite{Douezan2012}. The friction coefficients may be given as extensive or intensive quantities. In the latter case they must be multiplied by the surface area attributed to a node to obtain the nodal friction value to be used in the friction matrix.

The first and the 2nd terms on the rhs represent the in-plane passive elastic forces and bending forces in the cortex, respectively, and are both depending on the elastic properties of the cortex (elastic modulus $E_{cor}$, its Poisson's ration $\nu_{cor}$), and the cortex thickness $h_{cor}$. These forces are computed using the distance information between vertices and orientation angle of the triangles around with common hinge ( Fig.\ref{fig:cell_division}A).
The third term on the rhs represents the active cortical tension forces that the cell tries to maintain via cortex contractility \cite{Runser2024}.  The fourth and fifth terms,  $\vec{F}^{}_{vol,i}$ and $\vec{F}^{}_{osm,i}$ are forces due to volumetric changes of the cell and water in/outflow through osmosis that can be controlled by the cell, both resulting in a pressure exerted on the cell surface. The pressure can be computed considering the basic equation:
\begin{equation}
\label{eq:press_eq}
p =  - K_V\log \frac{V}{V_0}, 
\end{equation}
where $K_V$ is the bulk modulus reflecting how much it resists to compression, and $V$, $V_0$ are the actual and reference volume of the cell respectively. The above formula can be modified to include dissipative effects during cell compression or relaxation \cite{VanLiedekerke2019}. The cell volume can be easily computed using all nodal positions. Knowing the surface area of the triangles $A_T$, the pressure is then converted to nodal forces.

The terms $\{ \vec{F}_{adh,i} , \vec{F}^{}_{rep,i} \}$ describe the adhesion and repulsion forces on the nodes in presence of nearby objects such as another cell. Similar to the SEM, these can simply implemented as pairwise forces between nodes. However, this implementation may not prevent  that (significant) interpenetration of the surfaces of two cells occurs, particularly under high forces. The original model proposed by Odenthal et al. \cite{Odenthal2013} largely solved this problem. Here, the contact is resolved using triangle information to compute relative orientations and contact surface areas of for nearby triangles of the interacting cells. The geometrical information allows to apply a local discretization of continuum theory for adhesive bodies using physically tangible parameters such as the specific adhesion energy ($W$) and adhesive interaction range, which can be related to physical properties of cadherin bonds. A working simplification of this contact model is described in \cite{Cuvelier2023}. The results of pull-off simulations, in which two cells are separated, are consistent with those from analytical models of adhesive vesicles \cite{Smeets2019, BrochardWyart2003}. 
In cell-surface contacts in live cells, at the molecular level, cadherin bonds are constantly formed and broken, entailing dissipative effects. Hence adhesive contacts between cells induce shear resistance when moving past each other entailing an apparent friction between them. Observed cell-cell friction is likely a complex interplay of molecular dissipative effects and bond dynamics, with the latter depending on the type of adhesive bonds, their spreading density over the cell surface, and their typical lifetime. Stronger, long-lived bonds will likely result in an increased apparent friction between cells. In the DCM however, adhesive strength and pure viscous friction between nodes interactions are distinguished and they are parametrized independently of each other. This means that for example, in the  hypothetical case of a  cell adhered to a surface having zero viscous friction coefficients, the cell could move parallel to the surface without experiencing shear resistance. 
Nevertheless, when non-zero friction coefficients are assumed, those will interfere with the adhesion strength as more adhesion means larger cell-cell contact surface area and hence more fiction. In section \ref{sec:Monolayers and spheroids} we briefly demonstrate that adhesion strength influences with the viscous shear forces that cells experience when moving over a substrate.

The cell surface can show undulations due to thermal and active fluctuations, which can be modeled by a stochastic force $\vec{F}_{fluc,i}(t)$  \cite{VanLiedekerke2019}. 

When surrounded by ECM or attached to a substrate, cells may migrate. An straightforward way to model the movement of cells is by considering a global (biased) stochastic force term $\vec{F}_{mig}$ which can be distributed evenly on the nodes.  More fundamental approaches model consider the local formation of filopodia and the cytoskeleton (de-)polymerization during the migration process\cite{Kim2018, Tozluoglu2013, Chen2018, Heck2020,Vangheel2025}.

The linear system of equations \ref{eq:motion_DCM} can be written as concisely as
\begin{equation}
\label{eq:lin_system}
\Gamma \vec{v}  = \vec{F}, 
\end{equation}
where $\Gamma$ is a $3N \times 3N$ matrix, $N$ being the total number of nodes. 
This system can be solved efficiently for each $\vec{v}_i$ using a conjugate gradient (CG) solver. Subsequently an Euler scheme can be used to obtain the positions of the nodes. We refer to some recent works on how these components can be implemented (\cite{Odenthal2013, Smeets2019,VanLiedekerke2020, Runser2024, TorresSanchez2022}). It is important to note that the above mentioned model components are not unique and different approaches respecting the overall structure are possible. For example, a basic approach to model cortex viscoelasticity is using a spring-damper model such as the Kelvin-Voigt model where the spring stiffness can be related to the apparent elastic modulus of the cell cortex and the cortex thickness: $k= \frac{\sqrt{3}}{2} E_{cor}h_{cor}$. The approach in  ref.\cite{TorresSanchez2022} however, models the cell cortex elasticity starting from continuum elasticity theory for shells with known constitutive laws, and uses a virtual work description and finite element discretization to compute the nodal forces. This may be more accurate and appropriate when trying to understand cortical elasticity from a more fundamental point of view, but it has a higher computational cost.

\begin{table}
	\centering
	\begin{tabular}{l c c c c}	
	\hline
	 \textbf{Parameter} & \textbf{symbol} &  \textbf{unit} & \textbf{value} & \textbf{ref} \\
	  \hline
	    \hline
           \textbf{Deformable Cell Model} & \textbf{} &  \textbf{}  & \textbf{} & \textbf{} \\
            \hline
          Cortex Young's modulus  & $E_{cor}$ & $Pa$ & $1000$ & \cite{Brugues2010} \\
           Cortex thickness  & $h_{cor}$ & $nm$ & $200$ & \cite{Brugues2010} \\
          Cortex Poisson ratio & $\nu_{cor}$ &  -  & $0.5$ &  \cite{Tinevez2009} \\
           Cell  radius & $R_{cell} $ & $m$ & $6.10^{-6}$ & estimation  \\
         Cell nucleus size & $R_{nuc} $ & $m$ & $2.5.10^{-6}$ & estimation  \\
          Cell Bulk modulus & $K_V$ & $Pa$ & $2500$ & \cite{Hoehme2010}, \cite{Tinevez2009}\\
          Number of nodes/cell & $N$ &  & $160-2600$ & \cite{VanLiedekerke2020}\\
          Cell-cell adhesion energy  & $W_{cc}$ & $J/m^2$ & $10^{-5} - 10^{-3}$ & \cite{Hoehme2010,VanLiedekerke2022} \\
          Subtrate adhesion energy  & $W_{cs}$ & $J/m^2$ & $10^{-6} -10^{-4}$ & \cite{VanLiedekerke2022} \\
          Intra-cellular friction   & $\gamma_{int}$ & $Ns/m$ & $1\cdot10^{-4}$ & \cite{VanLiedekerke2022} \\
          Cell-cell friction & $\gamma_{ext} $ & $Ns/m^3$ & $5\cdot10^{10}$ & \cite{Galle2005,Buske2011,VanLiedekerke2022}\\
          Cell-ECM friction & $\gamma_{ECM} $ & $Ns/m^3$ & $10^{8}$ & \cite{Galle2005,VanLiedekerke2022} \\
          \hline
          
	\end{tabular}
	\caption{Nominal physical parameter values for the cell properties and the force models in a basic DCM. An (*) denotes parameter variability meaning that the individual cell parameters are picked from a Gaussian distribution with $\pm 10 \%$ on their mean value. See \cite{VanLiedekerke2020} for more explanation on this parameters}  
	\label{tab:parameters}
\end{table}

\subsection{Cell growth and division}

Similar to the implementation in a CBM, modeling the growth of the cell (the volume increase) in a DCM is relatively straightforward and only needs local adaption of viscoelastic network properties.  In a CBM, the cell division process can be modeled as a single event, in which two daughter cells are put next to each other, replacing the mother cell at an instant \cite{Galle2005}. Another possibility is to regard it more as a continuous process, by representing it as a dumbbell in which the spheres are slowly separating \cite{Drasdo1995}. The cell division process on the other hand, is more complex task in a DCM. 

Two major alternative approaches have been considered for the division process (Fig. \ref{fig:cell_division}B-C). In the first and simpler algorithm, during the growth stage, the mother cell inflates steadily and serves as a cocoon for two small adjacent daughter cells, which are instantaneously created next to each other at the beginning of the division stage. During simulation of the division process they "inflate" artificially fast. They are constantly in (an adhesive) contact with each other while still confined by the mother envelope. This inflation stops when the volume of the daughter cells equals that one of the mother. In the next step, the envelop of the mother cell is removed and the two daughter cells mechanically relax with the environment (e.g. neighboring cells). In this model, cytokinesis and other detailed biophysical processes in the mitosis phase are not modelled explicitly. 
The second approach  models the stages of cytokenesis continuously from the formation of a contractile ring leading to a furrow at the end of cell division process, followed by a complete separation of the cells and subsequent adhesion \cite{Bodenstein2020,Okuda2023, Cuvelier2023}. Due to the incorporation of this detail, this approach requires re-meshing procedures of the cell surface, entailing more complex algorithms and longer simulation times, but the approach may be considered if these phenomena are considered as important in the study.  

\subsection{Extensions: subcellular structures}

A basic DCM represents the cell surface and cortex, and approximates the cell interior by a homogeneous isotropic fluid. By a proper choice of parameters, the mechanical response of a cell upon mechanical forces applied to it can be quantitatively reproduced \cite{VanLiedekerke2020}.
However, the model naturally allows to incorporate the internal structures of the cell, such as a nucleus and structural cytoskeleton, hence making the model increasingly look as a "digital cell twin". The added detail allows to study more complex systems such as coupled mechano-chemical pathways (i.e. mechanotransduction) \cite{Mousavi2015}. In such models incorporating a proper nucleus element becomes key as the nucleus is a mechano-sensitive object \cite{Dupont2011} and known to have an effect on the mechanical properties of the cell itself. A nucleus may be approximated as a (rigid) object inside the cell or as a second deformable object \cite{Runser2024}. The internal cytoskeleton that transmits the forces between the cell boundaries and the nucleus can be included in the model by introducing additional elements connecting the nucleus with the cortex of the cell,  forming a coarse-grained approximation of these structures \cite{Kim2018,VanLiedekerke2020,Miotti2025}. Evidently these extensions increase computational cost as well as require calibration of extra model parameters.

\subsection{Extensions: surface mesh refinement}

The standard DCM, describing a deformable fluid-filled vessel, naturally allows to describe moderate to large deformations. A high degree of cell deformation may be anticipated for by attributing an increased number of vertices per cell to warrant accuracy \cite{VanLiedekerke2020}. Nevertheless, if the deformations become excessively large during the course of the simulation, errors due to deformed triangles may be induced and a (local) re-meshing may be necessary. The problem is reminiscent to re-meshing procedures in Finite Element Methods.  Re-meshing means that one creates or redistributes the edges of the vertices locally (introducing triangle splitting, merging or swapping operations). Mesh refinement operations may not be necessary in all situations. It may be imposed when the local curvature properties of the cell boundary starts exceeding the typical edge length in the triangulation. Such situations may arise when cells are heavily squeezed or actively deform (e.g. formation of filopodia). In those cases, the triangles can become distorted or degenerated such that the underlying models do not accurately capture the force distributions in the cell surface anymore. Re-meshing may also be necessary in case the elastic responses in the cell cortex are weak and fluid-like  behavior is observed (e.g. surface tension controlled fluid-like membranes, see \cite{Cuvelier2023,Vangheel2025}). In those situations the elastic forces that keep the vertices together are too weak causing them to easily diverge from each other, and triangles may become distorted quickly. However, re-meshing can be a tedious operation requiring attention with regard to the implementation. In particular, one must be sure that no extra energy or impulse is given to the vertices during re-meshing \cite{Cuvelier2023, Runser2024}. 

\subsection{Model calibration and validation}

High resolution models have more parameters to be calibrated than their simpler center-based counter-parts. Similarly to the CBM, parameters may be obtained from phenomenological laws or experiments (e.g. the cell cycle duration from tumor growth curves). An advantage of DCMs is that they can be trivially  deployed in a single cell experiments that affect cell shape (e.g. optical tweezers, optical stretchers, micro-pipetting, see Fig \ref{fig:single_cell}) or imaged-based techniques involving small cell numbers \cite{Vanslambrouck2024}. If shape changes relate to cell parameters, these parameters can then be inferred from the shape changes observed experimentally.

\begin{figure}[h!]
    \centering
    \includegraphics[width=1\linewidth]{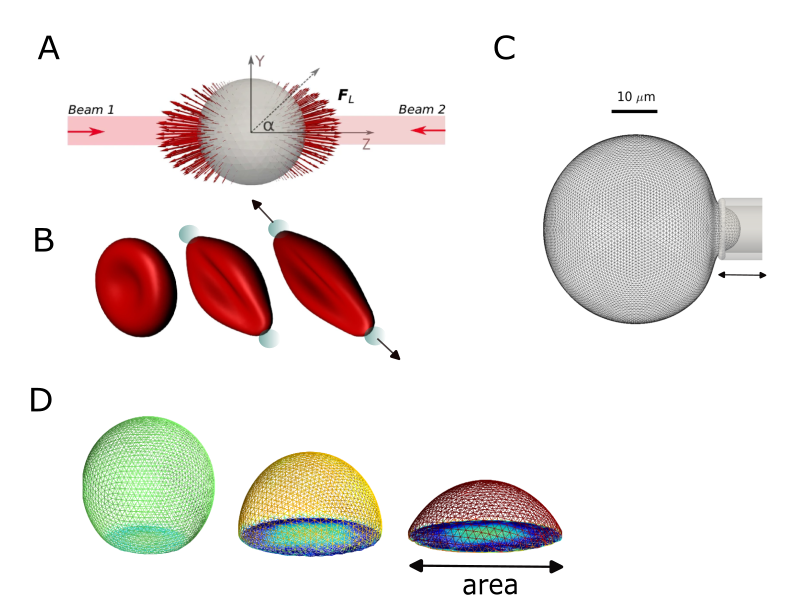}
    \caption{ Single cell experiments to calibrate a DCM. (A) optical stretcher.  (B) Optical tweezer. (C) Micro pipetting. (D) Cell-surface spreading experiment. Images are taken from \cite{Guyot2016, VanLiedekerke2020, Odenthal2013} and modified. }
    \label{fig:single_cell}
\end{figure}

Generally, biological data is often noisy and sparse. In many cases the model parameters can be only estimated by comparison of simulations with experimental data sets of low detail, for example covering only cellular-averaged variables.
A first milestone in model development is to perform a parameter identifiability (structural and practical) analysis which also involves a sensitivity analysis to identify the most influential parameters and those that potentially can be ignored and fixed at generic values or kept in a "physiological range". This can be achieved using Global Sensitivity Analysis, such Sobol' indices method \cite{Saltelli2008} or screening methods \cite{Morris1991}, the latter being cheaper in execution. Statistical techniques based on decision trees may also be used for ABM \cite{Retzlaff2023}.

The calibration of the parameters can be achieved by defining a multi-variate loss function that results in a least squares problem for which standard optimization algorithms can be applied. We note here that such loss function is usually (strongly) non-convex and thus may require appropriate techniques such as (meta)heuristics, multi-start methods or derivative-free techniques (e.g. \cite{Hansen2006, Kochenderfer2019}) in order to avoid that certain parameter values found only represent a local minimum.  

Once such the optimal parameter set is found, one should avoid to regard it as a single, fixed value. Due to the natural variability in cells such estimators should have a confidence interval. The width of such confidence intervals depend on the data availability and uncertainty thereof. The confidence over a parameter value is strongly related to its identifiability. Parameter confidence intervals can be estimated using "frequentist" statistical techniques based on Fisher Information or Profile likelihood \cite{Wieland2021}.  Bayesian techniques on the other hand,  generate a-posteriori distributions of the parameters rather than a confidence interval. Bayesian approaches in ABM have become increasingly popular in recent years \cite{Lima2021, Brinas2024,Visscher2024}. However, one should note that they may require many simulation runs (e.g. in a Markov Chain Monte Carlo algorithm) to obtain such distributions. This may be insurmountable when the model is expensive to evaluate, which is usually the case for a DCM.  In such cases, first building a surrogate model of the high fidelity model may bring a solution (see last section). The outcomes of Bayesian approaches may also depend on the prior information (i.e. prior distribution of the parameters).  
Finally, model validations should be performed to estimate the robustness and predictive power of the model. In principle calibration and validation should involve rigorous cross-validation procedures schemes that can quantify the expected model errors and uncertainties  \cite{Lima2021}. In practice however, data scarcity will often limit the validation.

\section{Computational efforts}

Compared to center-based variants, DCM simulations are significantly more expensive in terms of computing time due to the high number of degrees of freedom which needs to be integrated as most degrees of freedom are associated with vertices of the cell. The number of vertices per cell depends on the imposed system detail but typically there exist a lower bound for which below the simulation start to show artifacts and inaccurate behavior. While this needs to be verified, a finer surface discretization implies that there are more contact checks and interaction calculations per timestep, while simultaneously a lower timestep may be required to warrant numerical stability. As such with a DCM one can only target relatively small or scaled systems. A recent study \cite{Runser2024} nevertheless claims to reach up to 100,000 cells in a few hours of simulation time however at the cost of some compromises. There are a few "tricks" to speed-up the DCM simulations significantly. One possibility lies in the fact that using higher friction values (components of the friction matrix, see Eq. \ref{eq:lin_system} ) will stabilize the numerical scheme and allow larger timesteps. However, this may perturbate the intrinsic timescales in the system. Related to this, one may also artificially shorten the cell cycle time which will result in an apparent speed up in cell generated per simulation time. This may be justified for example if there is a clear separation in time scale between cell cycle time duration and the mechanical relaxation time of the cells.
Another possibility is to explicitly omit the node-node friction values in Eq. \ref{eq:motion_DCM} ) which is equivalent to a negligible friction between two cells. Technically, this will decrease the condition number of the friction matrix (\ref{eq:lin_system}) and allow faster convergence of the CG algorithm. However, this basically implies that shear friction in the tissue is ignored. We recall, however, that the strength of the DCM lies in representing the highly complex cellular systems accurately, rather than in being able to simulate very large systems.

Can simulations be speed up by parallelization? This depends on the research question. Simulating larger systems (more cells) becomes feasible as spatial parallelization schemes (e.g. by domain decomposition) can efficiently distribute computational load over High Performance Computer (HPC) platforms with low communication latency. Unfortunately, there is no such thing as "temporal domain decomposition", because time integration is a sequential process. In short, the spatial size can be increased by using more hardware, whereas the temporal reach of DCM simulations requires faster hardware. Hence, significant up-scaling for longer simulations (even assuming constant cell numbers) is less likely to be achieved in the future. Assuming that Moore's law will not be maintained in the coming years, one should probably focus on algorithmic advancements.
An alternative approach to "scale-up" simulations is to develop coarse-grained and hybrid models. In this context, the DCM has the advantage that it can be readily used in combination with a coarser scale off-lattice models such as the CBM \cite{VanLiedekerke2019,VanLiedekerke2020}, because
both DCM and CBM operate in the same continuous space and contact detection based on the triangles can deal with a large variety of geometric objects, including the basic spherical shapes.
This permits for example a workflow of calibrating and fine tuning low resolution CBM parameters such as the JKR contact force models from DCM simulations \cite{VanLiedekerke2019}. It also allows to create hybrid models, in which CBM and DCM operate simultaneously. In a recent work \cite{VanLiedekerke2022} a DCM is used in areas to resolve the phenomena requiring accurate cell shape while the CBM is used in the areas where no such accuracy or detail is required. 
However, up to date, the replacement of DCM simulations using equivalent center-based models remains a research area in development.

\section{Model and software availability}

A challenge for developing and using a DCM (up to date) is its availability in software packages. While the CBM has been developed for many years and is currently available in several software packages (e.g \cite{Hoehme2010, Ghaffarizadeh2018, Cooper2020, Breitwieser2022}), this is much more limited for the DCM. While the code architecture may be the same as for CBM (particle-based code architectures), new developments can require significant modifications, which may be restricted by the software architecture and the license of the software. Current open-source software packages for which the architecture allows to implement such model include:
BioDynamo \cite{Breitwieser2022}, TissueForge \cite{Sego2023}, Computix \cite{Pesek2024}, SimuCell3D \cite{Runser2024}, Edgebased \cite{Brown2021}, SEM++ \cite{Chattaraj2023}. 

\section{Applications}
\label{sec:Applications}
A DCM is well-suited to investigate mechanisms in cellular systems that require a high detail but involve only relatively low cell numbers. These are typically small individual systems (e.g. small spheroids, organoids, embryonic systems), or larger tissues that are comprised of smaller, representative and periodic sub-systems (for instance liver lobules \cite{Hoehme2010,Zhao2025}). In the following sections we discuss some multi-cellular systems in which a DCM can be particularly useful for investigation. We emphasize that the following simulations are more exemplary but merit to be studied more in-depth. In all the simulations we start from a basic DCM structure. In the simulations containing a nucleus, the latter is considered rigid and free-floating in the cytoplasm.

\subsection{Monolayers and spheroids}
\label{sec:Monolayers and spheroids} 

The growth dynamics of monolayers and spheroids have been extensively investigated in the past using center-based models, mostly predicting growth curves under various conditions of glucose/oxygen and other metabolite  concentrations or mechanical stress (e.g. \cite{Drasdo2005, Galle2005,Fletcher2010, Macklin2012, Poleszczuk2016, Gong2017, Jagiella2016,Lima2021}). DCM simulations (see \ref{fig:spheroid}A) can be used for this as well, yet targeting more fundamental mechanisms. DCM simulation can give accurate information on the role of cell deformation, interstitial spaces as well as specific properties such a cortical tension in spheroid mechanics \cite{Smeets2019,Ongenae2025,VanLiedekerke2019}, see e.g. Fig. \ref{fig:spheroid}B. For this information standard CBMs are inappropriate \cite{VanLiedekerke2019}.

\begin{figure}[h!]
    \centering
    \includegraphics[width=0.9\linewidth]{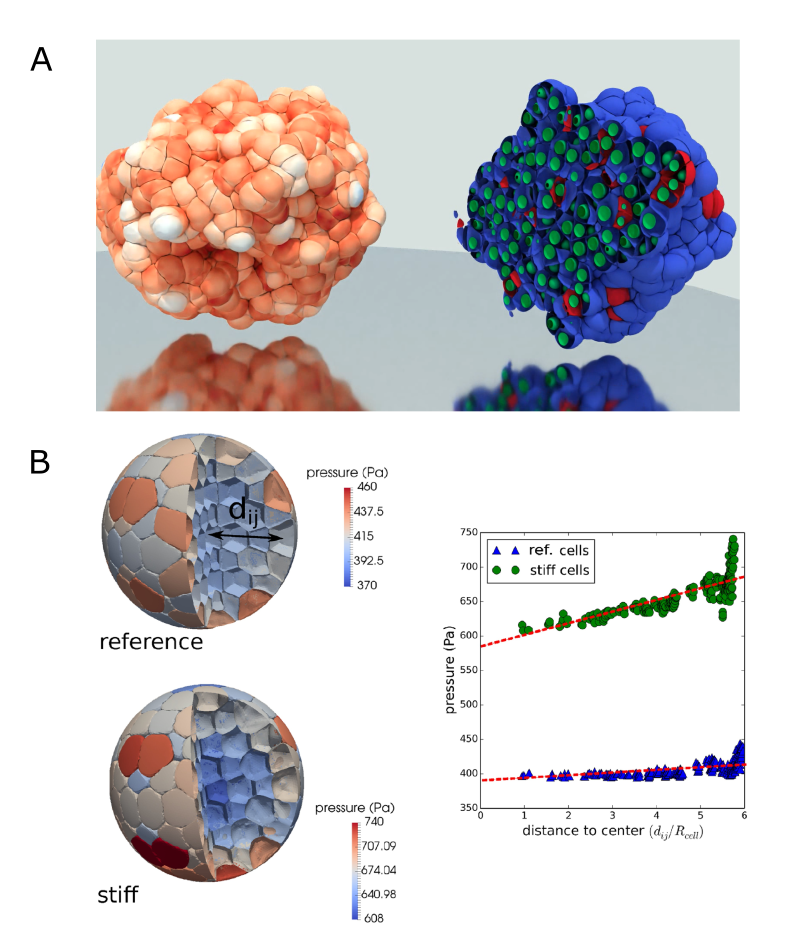}
    \caption{ (A) DCM simulation snapshot of a small growing spheroid with a cross section indicating the nuclei (right). (B) Simulations of an isotropic spheroid compression for two cell types with different cortex elastic properties, respectively with a normal (top) and stiff (higher $E_{cor}$) cortex (bottom). The color coding indicates the internal cell pressure. The right plot shows the appearance of a gradient in cell pressures along the spheroid radial distance in case of cells of the stiffer type, due to the fact that they  bear a larger part of the applied pressure from outside, thus "protecting" the inner cells. The softer type cannot do this as much, and hence  pressure is more equally distributed over the spheroid. Image taken from \cite{VanLiedekerke2019}}.
    \label{fig:spheroid}
\end{figure}

In monolayers, mechanical effects such as cells losing contact with the substrate can lead to cell death. In this particular effect, tracking precise cell shape may be critical and hence DCM simulations can bring more understanding of this process. As an example, we performed two monolayer growth simulations with a different cell type. For cell type 1, the adhesion to the substrate is moderate, whereas as for cell type 2, the cell-substrate adhesion is strong. This means that cells from type 2 will individually spread out more on the surface than type 1. We assume for simplicity that both cell types exert equal ( randomly and small) migration forces on the substrate. The simulations results, depicted in Fig. \ref{fig:monolayer}, show a clear distinction in the multi-cellular situation; for the cell type 1, the cells are more dispersed than for cell type 2, in which they are more crowded and occasionally are extruded from the monolayer. This on the first sight perhaps counter-intuitive result can be understood as a stronger cell-surface adhesion inhibits faster cell movement and hence relaxation of the pressure. The spread of the cell population is smaller for cell type 2 than for cell type 1, resulting in a higher pressure, eventually in pushing out of some cells of the monolayer plane. Despite the relative simple assumptions made in this system, this is a frequently observed phenomenon in epithelial layers \cite{Marinari2012}.

In summary, DCMs are well suited to study the effect of biophysical parameters on the fundamental mechanical processes in monolayers and spheroids, including for example cell expelling in monolayers, epithelial cell intercalation, growth under stress,  and spheroid surface topology. Currently less suited for simulating  systems with a size larger than  25 $  \mu m$ (spheroids) or 100 $\mu m$ (monolayers), as this implies too high computational cost.

\begin{figure}[h!]
    \centering
    \includegraphics[width=1.\linewidth]{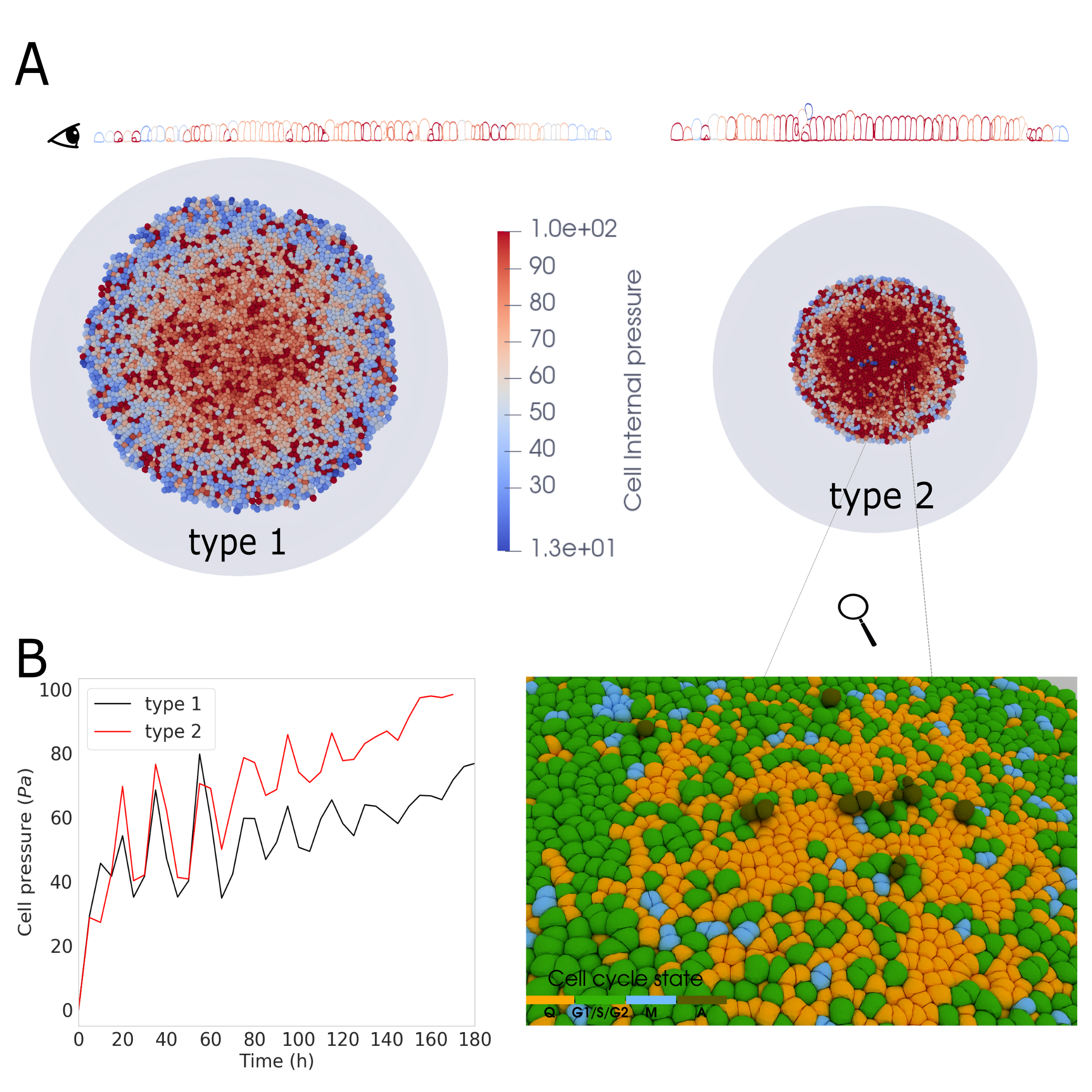}
    \caption {A: DCM monolayer simulation with respectively low (cell type 1, left) and high cell-substrate adhesion energy (cell type 2, right) at the same time point. Color coding is according to cell internal pressure, i.e. its state of being compressed. A vertical cross-section showing the cell shapes is given as well. Bottom: A zoom-in for the right case shows cells popping out of the monolayer and becoming apoptotic  (brown color) due to substrate contact loss. B: Evolution of average cell pressure for type 1 and type2. }
    \label{fig:monolayer}
\end{figure}

\subsubsection{Micro-carriers and micro-scaffolds}

Micro-scaffolds and micro-carriers are currently of high interest to deliver cells for targeted tissue engineering. The base material to fabricate them are bio-compatible polymers, yet materials can differ in their porosity, optical properties, presence of animal components, and chemical and physical surface properties \cite{Duan2023}. Understanding how cells grow on curved and complex surfaces is key for control of optimal cell yield or cell quality. A recent modelling study used a CBM to predict the growth of cells on smooth spherical micro particles circulating in bioreactors \cite{CantareroRivera2024}. However, due to the high curvature of such carriers, cell shape will likely play an important role in the growth dynamics and time scales to reach confluency (i.e. the part of the substrate surface area that is covered by cells). Here, we illustrate that a DCM approach can be used to further investigate this in detail. In our simulation, 6 cells are initially positioned on a non-porous carrier  (Fig. \ref{fig:carrier}, top). We consider a smooth carrier which is modeled by a triangulated surface.  During cell growth, we monitor the cell confluency in time (Fig. \ref{fig:carrier}, bottom right). Such simulations can give insights, for example in how fast 100\% confluency can expected to be reached as a function of carrier properties as well as cell properties, and whether predicting the moment that cells can be expelled from the carrier (Fig. \ref{fig:carrier}, right).

\begin{figure}[h!]
    \centering
    \includegraphics[width=0.8\linewidth]{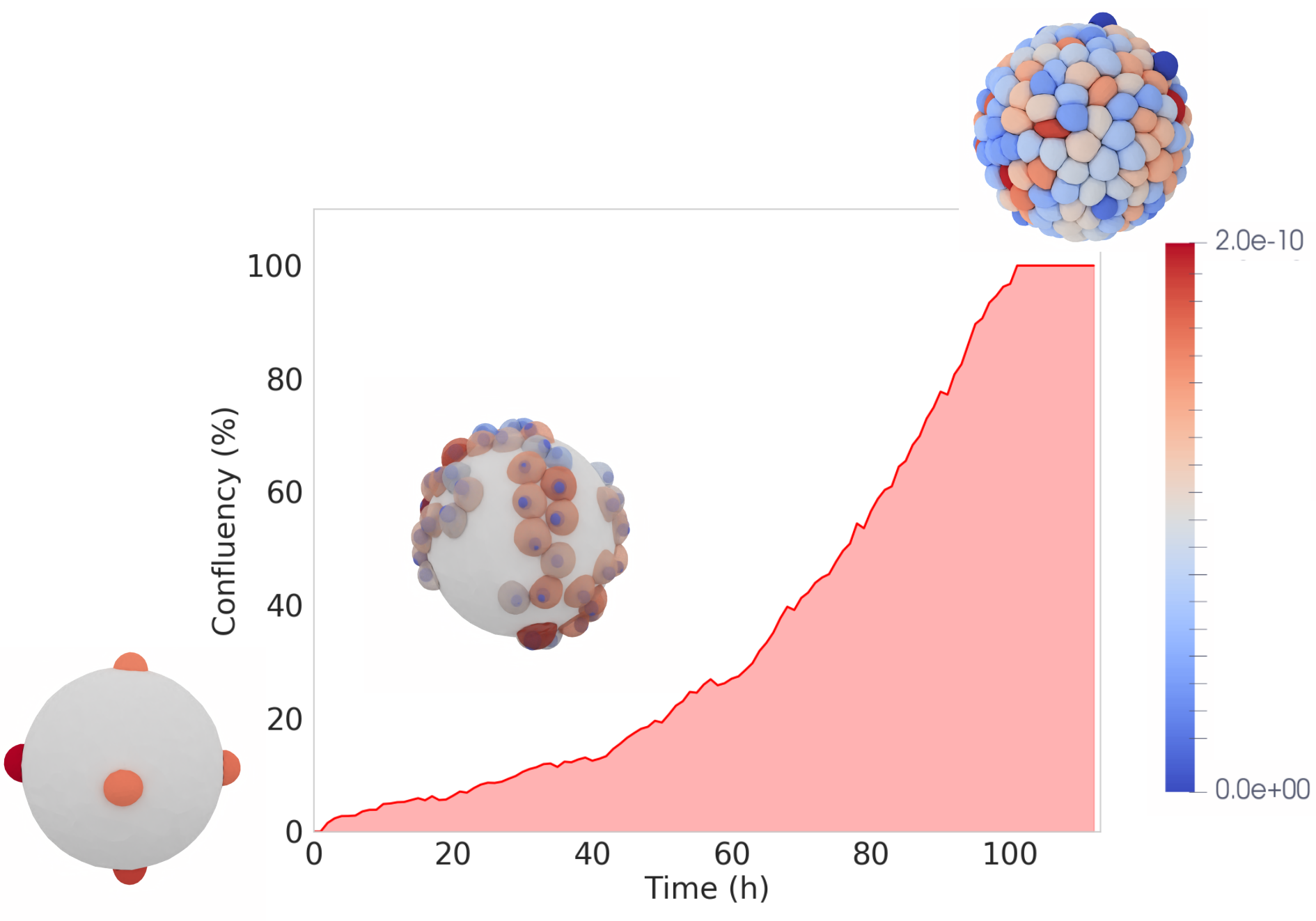}
    \caption{Micro carrier simulation with initial state (left) and plot of confluency over time  with coloring according to cell-carrier contact surface area.}
    \label{fig:carrier}
\end{figure}

Cells can also grow in scaffolding structures and it can be expected that scaffold properties (e.g. porosity) will likely affect the cell yield. For example, knowledge and prediction of the average contact area that cells have with the substrate can be of high interest if bio-active molecules are attached to the scaffold. To model this, we consider a system of cells positioned in a porous rectangular micro scaffold, modeled as a regular lattice of triangulated hollow cylinders (radius 1/4 of the cell size) that intersect each other. In total, this meshed geometry contains roughly 100,000 triangles (Fig. \ref{fig:scaffold}A). At the intersections of the cylinders, re-meshing operations (using CGAL, see \cite{VanLiedekerke2019}) are performed to smooth out irregularities.  Here, we performed two simulations, one in which the cells have a relatively high affinity with the scaffold substrate, and one in which this is low (i.e. the cells tend to adhere more to each other). The simulations start in each case from one cell. The plot shown in Fig. \ref{fig:scaffold}B shows the difference in cell-averaged average cell-scaffold adhesive contact area. As one may expect, the cells with a higher substrate affinity have more specific contact area compared to cells with relatively higher cell-cell adhesion. Another observation is that the former seem to likely to form a branched-like morphology, wheres the latter are more inclined to acquire a spheroid-like morphology within the scaffold (\ref{fig:scaffold}C-D). Nevertheless, the evolution over time may be non-trivial. We note here again that active migration processes, which may alter these results, are not included in these simulations.

As we demonstrated, a DCM is capable of mimicking sophisticated experimental culture systems including microscaffolds and microcarriers to great detail and study interactions of cell biomechanical and geometrical properties. The triangulation-based discretization, contact detection, and interaction of the DCM can naturally deal with most complex geometries, usually readily available as STL files.  
This permits to study in how far observations can be associated to purely physical effects opposed to active cellular control for a given culture systems, and allows to quantify the effect of geometry between complex culturing systems if the cells in both systems were the same.

\begin{figure}[h]
    \centering
    \includegraphics[width=1\linewidth]{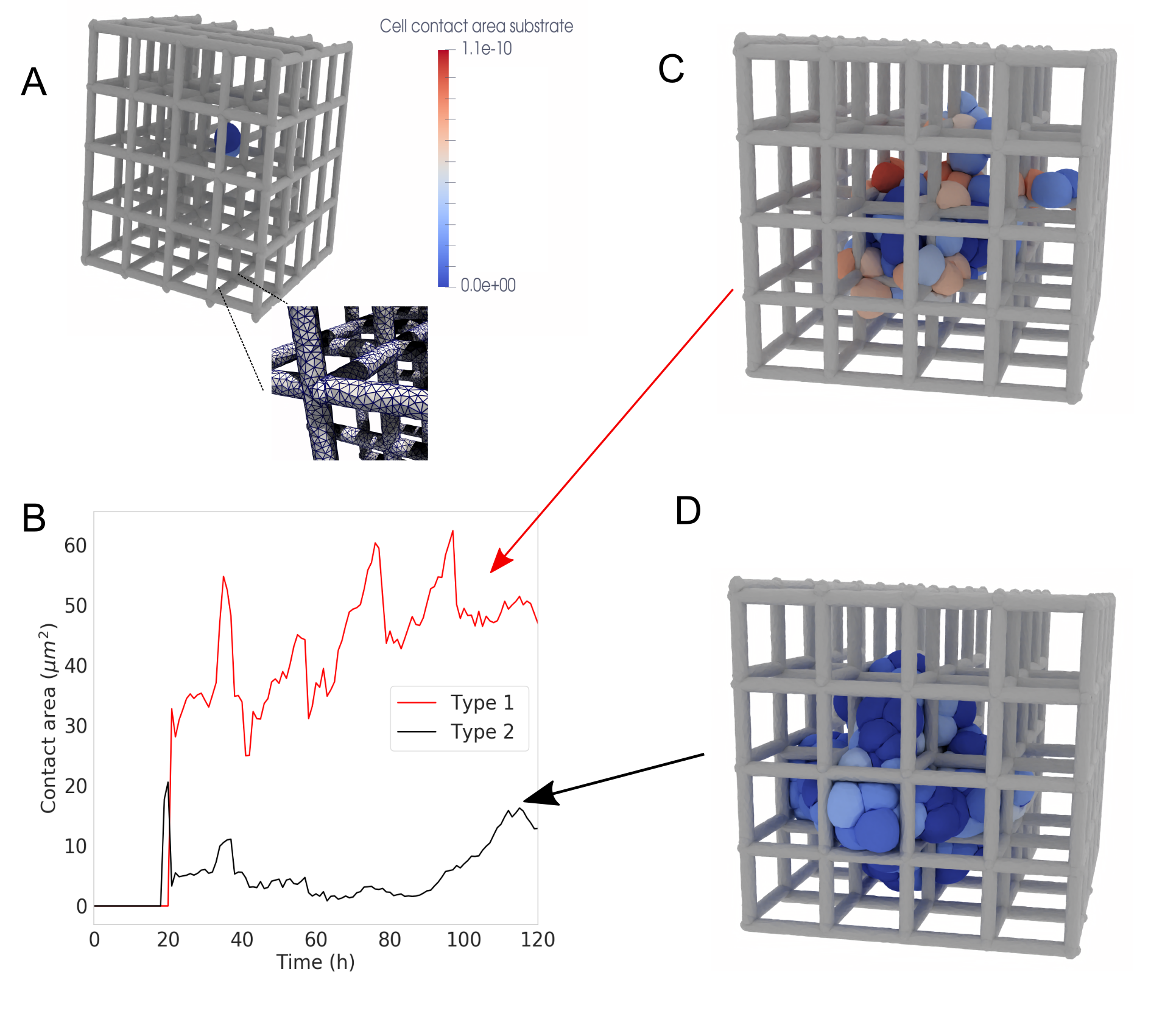}
    \caption{A: Initial situation: triangulated micro-scafold model with 1 DCM inside. B: Time plot of the average cell-scaffold contact area for two cell phenotypes with different cell-substrate adhesion energy (red : high cell-substrate affinity; black: low cell-substrate affinity). C-D: Final state of the simulations for the two cell types. Coloring is according to cell-carrier surface area.}
    \label{fig:scaffold}
\end{figure}

\subsection{Organoid-like structures}
\label{sec:Organoids}

Organoids are tissue-engineered cell-based in vitro models that recapitulate many aspects of the complex structure and function of the corresponding in vivo tissue. They help to understand the self organization of cells and the relation to tissue morphology, and will have potentially a large impact on medical therapies. It remains a challenge to standardize organoid cultures for experimental manipulation that would lead to a maximum benefit for understanding and controlling human health and disease. Computational models (e.g. \cite{Hannezo2014,MontesOlivas2019,Laussu2025,Zhang2025}) facilitate studying the multiscale dynamics of organoids in addition to experiments. Here we show how the DCM can explain the dynamics of a small system of bio-engineered cells (a few thousand cells). We mimic the growth of stem cell based cysts for a particular technique in which cells are encapsulated by an alginate spherical capsule \cite{Cohen2023}.

\begin{figure}[h]
\begin{center}
\includegraphics[width=4.2in, height=4.in]{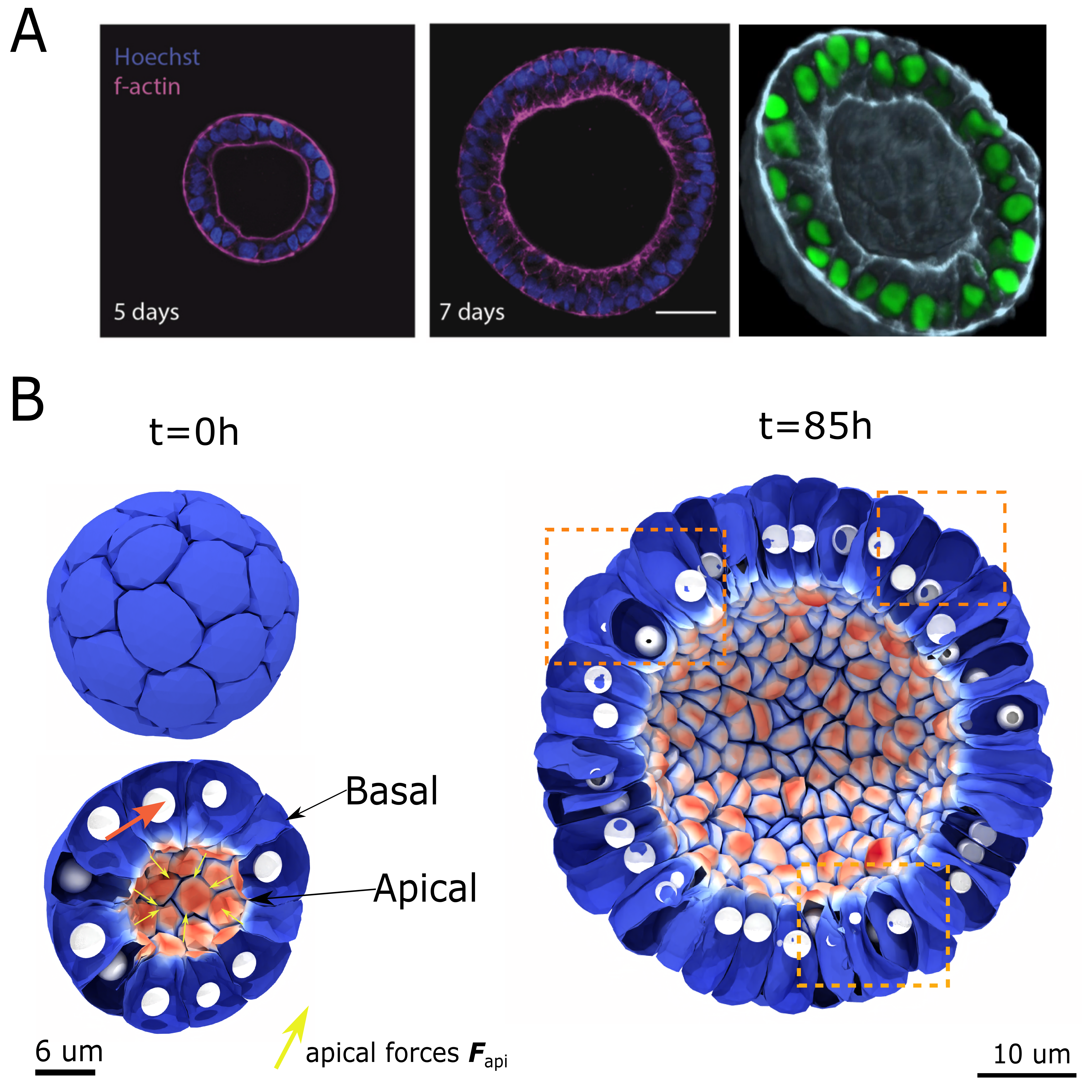} 
\end{center}
\caption{{ (A) microscopic images of an stem cell organoid (cyst). (B) Simulation of an organoid starting (t=0) from 42 cells positioned on a sphere. The yellow arrows denote the forces $F_{api,i}$. The red arrow indicates a cell division direction. Right: organoid simulation at t=85h. The dashed frames indicate presence of pseudo stratification. }
  }
\label{fig:organoids1}
\end{figure}

In the experiments, in a first stage a small clump of cells assemble to an organised system with apical-basal polarity which develops a small lumen. The system then further grows while keeping a one-cell-thick layer around the lumen,  resulting in a cyst (Fig. \ref{fig:organoids1}A). This initial phase is mimicked by a system of 40 adjacent deformable cells which are positioned on a spherical surface, see Fig. \ref{fig:organoids1}B, left. Every cell consists of 162 computational nodes.  Starting from this stage, we let the cells grow, while keeping track of their apico-basal polarity vector which points toward the centre of the cyst. The cells all have a normal cell cycle and during division, the division vector (the normal to the division plane) is perpendicular to the apico-basal direction and tangential to the cell layer but otherwise (within the plane of the cell layer) assumed to be randomly oriented (Fig. \ref{fig:organoids1}B). This division orientation has been clearly observed in microscopic images of growing organoids \cite{Cohen2023,Ragkousi2014, Kara2018}.
This orientated cell division is likely a necessary condition to maintain a one-cell-thick layer demarcating the lumen in the beginning stage and later organoid stability. In the 3D model, simulations with pure randomly oriented cell divisions quickly result in a un-ordered system of cells not in agreement with the observed structure of later stage organoids (Fig. \ref{fig:organoids1}A, right). However, oriented cell division alone is likely not sufficient to maintain a one-cell-thick layer. This line of argument is supported by simulations of  blastula formation \cite{Drasdo2000} as well as 2D simulations of the intestinal crypt dynamic \cite{Drasdo2001} using a CBM  where a one-cell-thick layer could be maintained only if the cell division direction was oriented tangentially to the one-cell-thick layer and cells were keeping a certain  polarization.  

Here, as a stabilizing we consider two alternative mechanisms, which may work in synergy to each other. First, we assume the presence of a liquid inside the lumen exerting a pressure $P_l$ (due to osmotic balance) on the apical side of the cells. To simulate the effect of the lumen pressure $P_l$, forces are exerted to all nodes that border the lumen side \cite{VanLiedekerke2022}. A second stabilizing effect emerges from cell polarity effects similar as for the above mentioned blastula formation and crypt shape maintenance: From observations \cite{Ragkousi2014, Kara2018} there is evidence that cell nuclei and a part of the cytoplasm move away and towards the basal side (a process called interkinetic nuclear migration) during the cell cycle, while parts of the cell surface tend to keep contact with both the lumen localised at the apical side as well as to the basal side. We include the phenomenon that the cell keeps contact with the apical side in the model as follows (the same reasoning is applied for the basal side): (i) We identify the apical region of the cells, which are the triangles (or nodes) that border on the lumen (see \ref{fig:organoids1}B). (ii) We calculate a mean radius (apical distance) from all nodes that border on the lumen. (iii) The apical region has a dense presence of actin filaments keeping the cells tightly together. This effect is mimicked by introducing an extra force in the model: If a cell starts to either move away or move towards the lumen due to other forces, restoring forces $F_{api,i}$ appear to counteract this and try to keep those cells near the lumen.  Importantly, these forces will not oppose the global displacement of all cells during organoid and lumen growth on longer timescales. We assume that both viscous and elastic effects contribute to the nodal forces:
\begin{equation}
    \vec{F}_{api,i} = - \mu_{api} \vec{v}_i - k_{api} \vec{d}_i.
\end{equation}
The first term $\mu_{api}$ is a parameter that controls the extra viscous friction in the apical ring. The second, elastic term is controlled by the distance $||\vec{d}_i||$ which is the deviation of the node from the mean apical spherical surface.

\begin{figure}[h!]
\begin{center}
\includegraphics[width=4.in, height=6.in]{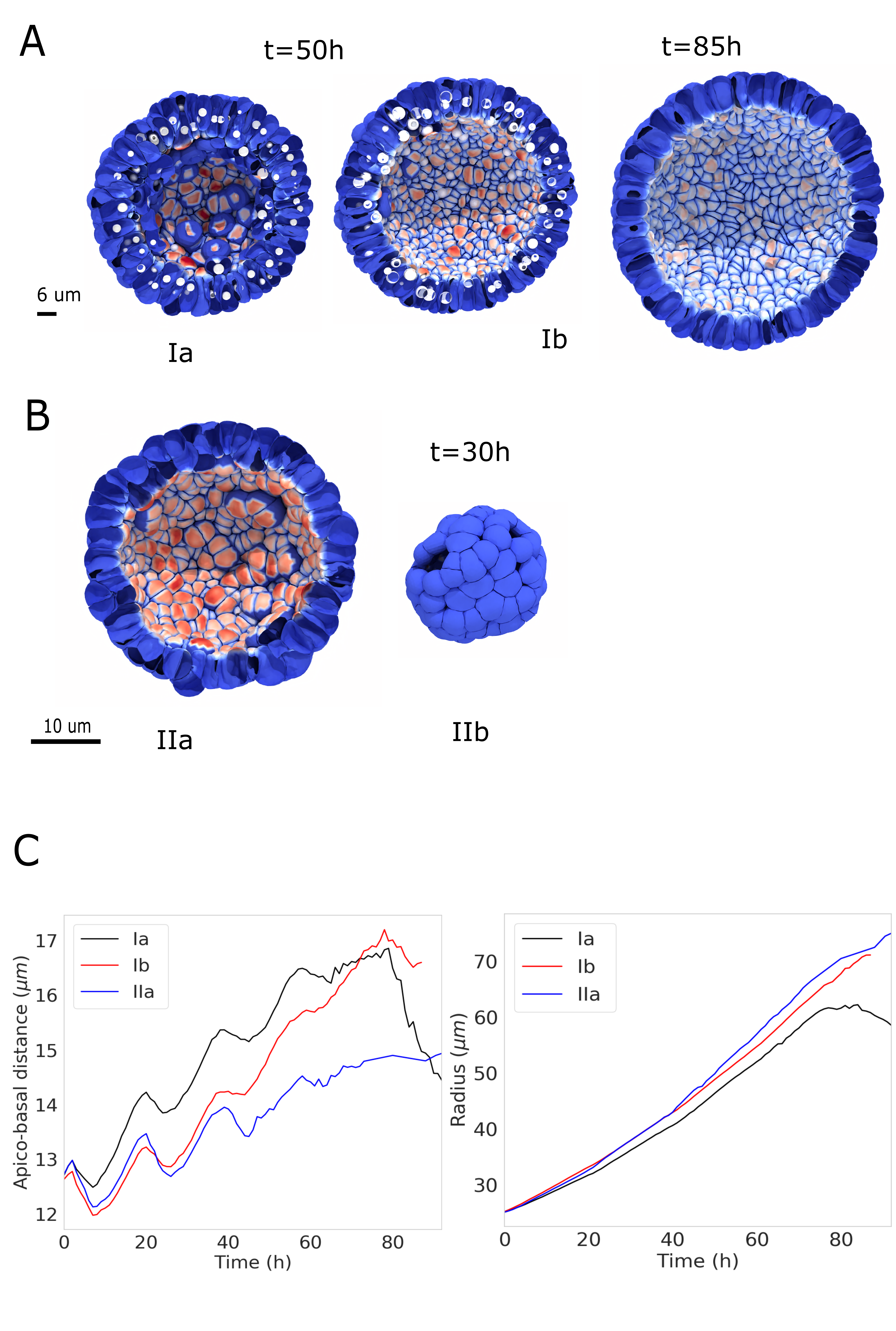} 
\end{center}
\caption{Organoid growth under different physical conditions. A: case Ia (high apical cohesion, no lumen pressure) and Ib  (high apical cohesion, with lumen pressure $P_l=25 Pa$). B: case IIa (low apical cohesion, no lumen pressure) and case IIb (high apical cohesion,  lumen pressure $P_l=50 Pa$. C, left: average apical-basal distance evolution in time. C, right: average organoid radius evolution in time.   }
\label{fig:organoids2}
\end{figure}

In addition, as in the experiment the cells are surrounded by Matrigel, we account for the presence of ECM at the basal side of the system. Again, we assume that this ECM behaves as liquid-like on long time scales, and does not prevent the expansion of the organoid. This can be quantified by an extra contribution of cell-ECM friction term in Eq. \ref{eq:motion_DCM} for the basal nodes.   
We performed a few simulation cases to mimic the effect of the parameters $\mu_{api}$, $k_{api}$ and $P_l$. All other parameters remain the same.
We let the organoid grow from its beginning configuration up to about 3-4 days.  We considered 4 different cases. In case (Ia) we assumed a strong coherence to the apical ring but the osmotic activity by the cells is regulated such that the lumen pressure remains low. In the second case (Ib) we implemented the same conditions but the lumen pressure is high. Both systems result in a spherical expansion. However in the case (Ia), the cells quickly start to show incoherencies as some cells "escape" towards the lumen side. This incoherence is a result of on an excess of mechanical forces and limited space to move for the cells and which is at the origin of a phenomenon called pseudo stratification, often observed in these systems. Case Ia also results in a limitation in size growth of the cyst (\ref{fig:organoids2}C, right). In case Ib on the other side, the system expands more or less smoothly keeping almost a single layer of cells even at later stages (t=85h), see Figure \ref{fig:organoids2}A. This results in an increased apico-basal distance of the cells (\ref{fig:organoids2}C, left) and faster growth of the organoid (\ref{fig:organoids2}C, right). We attribute this difference to the fact that the lumen pressure pushes the cells outwards, relaxing compressive stress generated by cell growth  and thus creating more space for dividing cells. In the low pressure case, the system becomes too crowded and the apical restoring forces on the cells cannot cope with the other (compressive) forces leading to local undulations. 
Local undulations or bulging can generally occur if the stabilizing forces cannot balance the destabilizing effects anymore \cite{Drasdo2000b,Drozdowski2024}. It has been observed in models that growth and division are destabilizing the spherical shape of a shell by rapidly  and locally generating compressive stresses which cannot be balanced by stabilizing elastic or bending resistance inside the shell. This may result in tissue folding  which is a key process in embryonic developement \cite{Trushko2020}. The occurrence of the effect further also depends on the radius of the hollow sphere \cite{Drasdo2000b}. Here, the formation of (pseudo) multi-layers i.e., cells positioned in a kind of staggered way in radial direction, is perhaps a response of the system to mitigate compressive stresses and avoid folding. In this process, cell-cell shear forces (supported by cell-cell adhesion) and viscous friction play a role as well, as for example high friction in the cell movement disfavors rapid stress relaxation. These factors can be investigated more extensively with the DCM.

Case (IIa) and case (IIb) represent systems where the apical ring strength and viscosity is a factor of 10 times lower than those of cases I, in combination a low and high lumen pressure respectively. Here we clearly observe that in case Ia, local undulations appear rapidly, with cells moving outside the apical and basal surface. However, if the lumen pressure is very high, cells are pushed too much outwards so that  other instabilities occur (Figure \ref{fig:organoids2}B). 
Here the high lumen pressure generates tensile stresses that cannot be balanced anymore by cell-cell adhesion or the generation of extra surface by cell growth,  leading to local rupture of cell-cell contacts. 
The rupture of cell contacts then causes fluid to exit the lumen, which in turn relaxes the lumen pressure so that the cell-cell contacts can re-form, which is why empty spaces may not be observed in reality. 

Concluding, organoids feature the combination of tissue complexity, three dimensionality and relatively small size. The above simulations show that the basic DCM with some extensions is a natural choice to study the biophysical origins of various particular aspects in organoid morphology, including apico-basal polarization, pressure control in fluid-filled lumens, and crypt formation. The DCM permits systematic studies of how properties at the individual cell level affect the morphotype of the three-dimensional multicellular structure and hence qualifies to better understand the underlying mechanisms of organoid formation and guide experimental design.

\subsection{Integrating mechanisms of early embryonic development}

Future bioengineering of organs and organ replacement tissue is often inspired by early mamalian embryogenesis. Computational modelling of  multi-cellular systems that include the coupling of mechano-biology with biochemical signalling can help to understand the underlying mechanisms \cite{Cockerell2023} and may then be used to design and engineer biological tissues with certain targeted properties. Although approaches with center-based models have been proposed (e.g. \cite{Nielsen2020,Drasdo2000}), because variations in cell shape are so prominent during the first stages of cellular organization, numerous researchers have used vertex based models (e.g. \cite{Fletcher2014, Okuda2018, Rozman2020, Jiang2023}). Yet, the limited shape adaption in VM may be a limiting factor in several cases \cite{Runser2024}.  A DCM approach allows to simulate phenomena at smaller spatial scales than VM, potentially giving a more detailed view on ongoing biological processes which are necessary to explain the above lying scales \cite{Bodenstein2020,Dokmegang2021, Dries2025,Cuvelier2023,Vanslambrouck2024}.

As an example, we here focus on a particular stage in lumen formation which is a precursor of bile canaliculi formation in liver. Understanding this process is key for the creation of functional liver tissue replacement systems for both liver toxicity testing as well as for bioengineering of liver tissue for prospective implantation. 
Bile canaliculi are thin tubes between adjacent hepatocytes that collect the bile secreted by the latter. They progressively form networks and eventually drain their content in the bile ducts from where the bile is transported out of the liver to the gal bladder \cite{Ober2018,Vartak2021,Vartak2021b}. Alterations in the orchestration of the bile transport system can be associated with severe diseases. 
Bile canaliculi are assumed to be formed by the micro villi-lined vesicles that are present in adjacent hepatocytes and progressively move towards the hepatocyte interface where they fuse into one large vesicle \cite{Chiu1990}. Subsequently the small lumen that are formed in each hepatocyte pair  merge into a tube-like structure \cite{Sigurbjornsdottir2014}. The stability of this lumen is assumed to be maintained by a mechanical equilibrium balancing osmotic pressure forces in the vesicle on the one hand, and the opposing forces originating from the tight junctions which seal the canalicular lumen, the local actin-myosin contractility, and the background pressure of neighboring cells on the other. Although high resolution microscopy and theoretical modeling has helped in describing and observing lumen dynamics \cite{Dasgupta2018}, a quantitative and mechanistic description of the physical variables and conditions necessary to lead from the initial vesicles into an organized tubular structure, is not yet given. 

Here we shed some light in how agent-based models may contribute to this.  As a proof of concept, we show here that the DCM can simulate the onset of bile canaliculus lumen formation as we can exactly describe the forces that come into play at this detail. In particular, we into look how osmotic activity and cell-cell adhesive tension influences the lumen dynamics.

\begin{figure}[h!]
\begin{center}
\includegraphics[width=5.1in, height=5.1in]{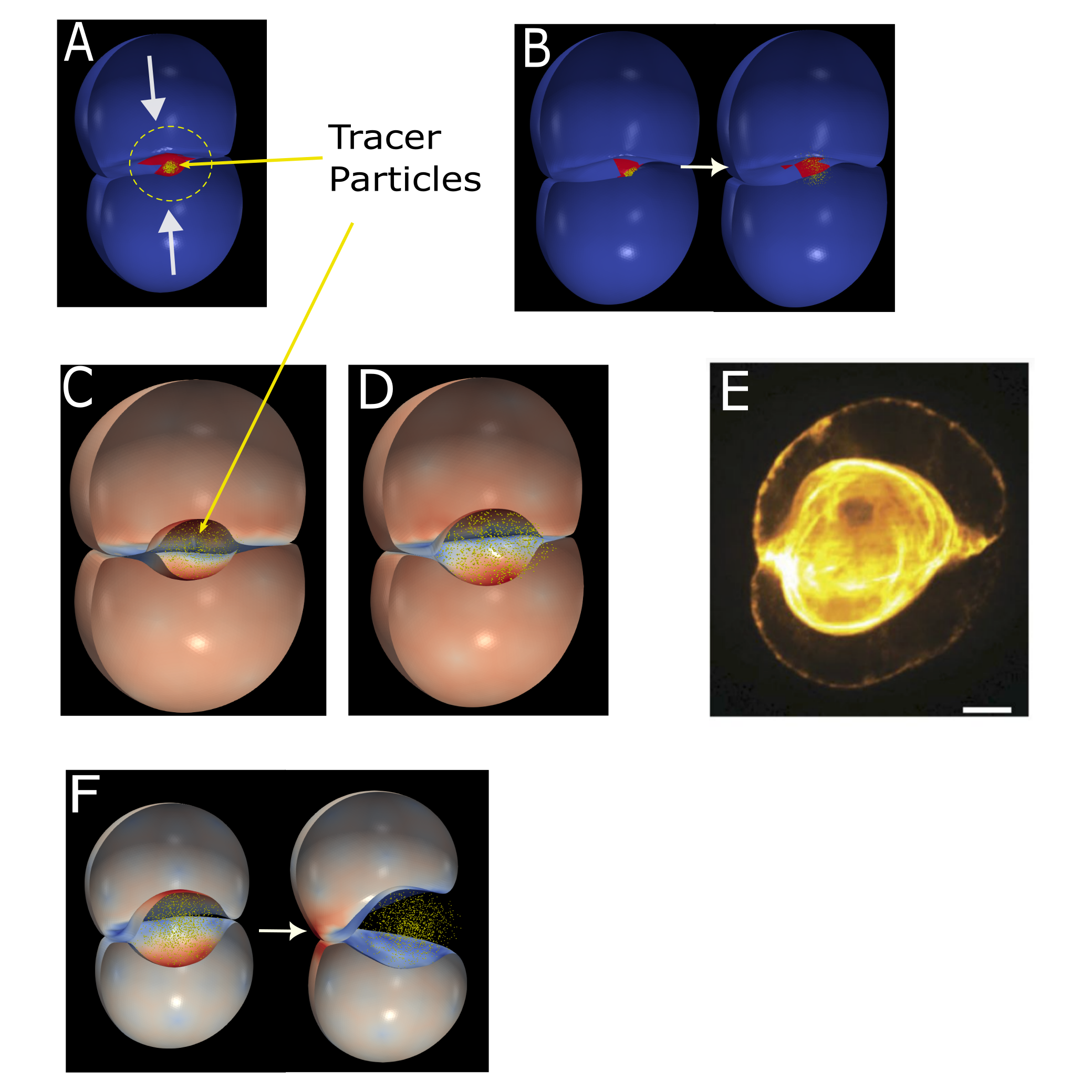} 
\end{center}
\caption{{\bf }
Snapshots of initial bile canaliculus formation. A: Initial cavity, the red colored triangles mark the cavity area, the arrows the cell polarization.The yellow particles are the tracer particles. B: The cavity collapses at low osmotic pressure ($W/P_L = 10^{-6}m$,). C : Equilibrium status at moderate osmotic pressure, ($W/P_d \sim  0.5\cdot10^{-6} m$ and $W/P_L \sim 0.4\cdot10^{-6}  m$ ). D: Experimental image of two cells with lumen, image taken from  \cite{Dasgupta2018}.  F: Simulation of high osmotic pressure ($W/P_L \sim  0.25\cdot10^{-6}  m $) leads to rupture of the cavity. The coloring marks the cortical tension of the cells.  }
\label{fig:fig11}
\end{figure}

We consider a simulation with two adhering hepatocytes which are polarized (see Fig~\ref{fig:fig11}A). A typical canalicular diameter varies from 0.5 to 1.0 µm. This implies we require a modeling technique with a subcellular resolution, hence we increase the nodal resolution of the DCM by a factor of 4 times  compared to the previous simulations above, resulting in approximately 2600 nodes per cell.
We assume that some of the released vesicles have already formed a small lumen between the adjecent hepatocytes. This lumen is represented only by a few triangles.
To model the local flux of salts that cause the osmotic effect, we introduce tracer particles \cite{VanLiedekerke2022} which are released for both cells in the lumen at the cell-cell contact in direction of the apico-basal polarization vector.  Those particles diffuse and can hit and mark the local cell surface triangles, upon which a pressure force $F \sim P_LA_T$ will be applied to them. Released particles by a cell cannot re-penetrate a cell surface and are trapped in the cavity. As such the cavity boundary is continuously and automatically marked. The osmotic pressure may then further widen the cavity.
The resulting lumen size dynamics is the result of an interplay between the osmotic pressure and the adhesive forces at the cell-cell interface, in line with ref. \cite{Dasgupta2018}. We now simulate three simple cases. First, we simulate the case where the cell-cell adhesion energy is high compared to the osmotic pressure (we choose here an arbitrary ratio  $W/P_L = 10^{-6}m$). For this case, the simulations show that the initial lumen gradually shrinks and disappears, see (see Fig~\ref{fig:fig11}B). In the second case, where the ratio of adhesion energy to osmotic pressure is lower (ratio  $W/P_L \sim 5\cdot10^{-7} m$ and $W/P_d \sim 4\cdot10^{-7} m$ ), a stable vesicle can exist, see Fig~\ref{fig:fig11}C-D). This conformation shows good resemblance to experimental images of cavities between two cells \cite{Dasgupta2018}. Finally for the last case the ratio is small (ratio  $W/P_L  \sim 2.5\cdot 10^{-7} m$), see (see Fig~\ref{fig:fig11}F. In this case the simulations predict that a stable vesicle is not be possible as the cell-cell contact can locally rupture resulting in the bile leaking into the inter-cellular spaces.  
The question remains now, how a such small cavities can aggregate into a mature multi-cellular lumen \cite{Guoa2024}. In the future, one could include above mentioned fine scale mechanism in simulations of larger systems of adjacent cells to explore the conditions necessary for these inter-facial vesicles to merge into mature lumens or tubular structures as visible in early tissue development. Recent work has already provided interesting hypotheses  \cite{Maarten2023,Belicova2024}. 

Concluding, a fundamental advantage of the DCM is that the resolution can straightforwardly be increased, thereby allowing to study the origins and development of smaller sized structures such as bile canaliculi. In addition, the particle-based nature of the model naturally permits to incorporate diffusive tracer particles that can used to mark particular areas in the tissue and mimic physical effects such as osmosis.

\section{Summary and outlook}

We started this article by giving an overview of agent-based high-resolution models that permit to represent realistic cell shape and mechanical behavior. The Deformable Cell model (DCM) has the potential of representing cells with a high fidelity of biophysical characteristics in a many biological questions and biotech applications. Altogether, the simulations with this model, although here only shortly and conceptually explored, demonstrate their potential by giving insight and quantifying the contributions of physical mechanisms at the cellular level and how varying them may affect the tissue scales. 
The examples included growth of monolayers on flat surfaces and spherical micro-carriers, as well as multicellular spheroids and cell populations in growing in scaffolds. We also illustrated how the DCM can provide understanding in the stages and processes during organoid growth. We simulated the early stage formation of cysts and the bile canaliculi formation in hepatocytes, which are of great importance in basic research (i.e. developmental biology) as well as applications in regenerative medicine.

Nevertheless, it must be stressed that in the formation process of tissues, accounting for mechanical effects and basic cell and tissue physiology alone is not sufficient to grasp the system complexity as cell signaling and metabolism pathways must be further explored and integrated too (\cite{Dichamp2023,Thalheim2022}). By construction agent-based models can combine physical models of cells with intracellular state models (e.g. cell regulatory or metabolic networks) \cite{Ponce-de-Leon2022} and extracellular models that inform about concentrations of molecules around the cells. As such they are particularly well-suited for studying communities of interacting cells. 
Generally, ABMs naturally allow the integration of cellular phenotypes estimated from genomics data \cite{Johnson2025}, whereby intracellular networks can directly be constructed from -omics data \cite{Montagud2022}. The ongoing developments in spatial genomics and tools permitting to integrate them into models (either in a rule-based or data-driven fashion) \cite{Retzlaff2023} will increasingly allow to improve ABM intracellular state models. As such ABMs will be transformed to hybrid knowledge-based and data-driven methods for deciphering the interplay of bio-physical parameters, spatio-temporal information, and biological processes. In this paradigm a DCM can become a specific powerful tool for understanding how subcellular scale processes affect formation, growth and homeostasis of (bio-engineered) multi-cellular systems.

For models serving as a digital twin of bioengineered systems, one may aim at simulation times below those of theexperimental execution times.
Because "high fidelity" spatial models (such as a DCM) are usually expensive execution time, especially for big cell numbers, they may be inherently too slow to be considered as a DT in that aforementioned sense. At this point, it can be necessary to build a surrogate or reduced model of the DCM. A surrogate model is a fast-to-execute variant that is trained on the results of the high fidelity model, at the cost of a reduced information output. Surrogate models can be based in several techniques. They can be physics-based, for example by using coarse-grained ABMs \cite{Lima2021}, or based on a wide set of machine learning techniques such as random forests, Gaussian processes or neural networks\cite{Jorgensen2022}. Especially the latter are fast to execute but may data demanding during training. The construction of a surrogate model  may be a long process itself, yet once such a model is available, it can make the model calibration and generate model predictions much faster.

All these ideas and developments, however promising, do need to be considered with necessary caution. The integration of model components in a working software package is a complex exercise that requires extensive coordination among researchers in multiple fields. Pursuing a wider acceptance and use of ABM in a broader community, we recommend to put a strong attention to following points:

\begin{itemize}
   
    \item Employing more standardization of algorithms in ABM codes. This is not always possible among different methods as they are rooted on different concepts (see e.g. comparison DCM, CPM, PFM, SEM, VM). 
     \item Agreement on unit tests to ensure that for well defined standard cases, different models provide the same verifiable answer. These tests can be based on biological experimental datasets or on model-wise verification, e.g. testing model results against analytical solutions or physical principles.
    \item Provide sufficient code availability (ideally, open-source and community based with long term support).
      \item Development of user-friendly interfaces that allow non-technical users to setup and run established  and tested models, to facilitate future usage in medical or biotech companies.
\end{itemize}

Some of these points are currently being addressed in recent ideas and initiatives \cite{Osborne2017,Cogno2024,Ntiniakou2025}.

\section*{Acknowledgments}
We would like to thank Pierre Nassoy (Université de Bordeaux) for all insightful discussions about cyst growth.

\section*{Author contributions}
Conceptualization: PVL, DD, KA
Article writing: PVL, DD
Algorithms and model development: PVL, JP
Software: PVL, JP
Financial support: KA

\section*{Software}
All simulation results have been generated using a modified version of(\href{https://multicellular_modelling.gitlabpages.inria.fr/group_webpage/tisim}{TiSim}). The code or executable (via docker image) will be mad available.

\newpage
\bibliographystyle{unsrt}

\end{document}